\documentclass[a4paper]{article}

\usepackage{a4wide}
\usepackage[utf8]{inputenc}
\usepackage[T1]{fontenc}
\usepackage[english]{babel}
\usepackage[bookmarks=false,hidelinks]{hyperref} 
\usepackage{lmodern,newunicodechar}
\usepackage[
  backend=biber,
  url=false,
  style=trad-plain,
  maxnames=4,
  minnames=3,
  maxbibnames=99,
  giveninits,
  uniquename=init]{biblatex}
\usepackage{microtype}
\usepackage{graphicx}
\usepackage{url}
\usepackage{import}
\usepackage{layout}
\usepackage[table]{xcolor}
\usepackage{amssymb}
\usepackage{amsmath}
\usepackage{tikz}
\usepackage{tikz-uml}
\usetikzlibrary{positioning,calc,shapes,fit,shadows.blur}
\pgfdeclarelayer{background}
\pgfdeclarelayer{foreground}
\pgfsetlayers{background,main,foreground}
\usepackage{comment}
\usepackage[group-minimum-digits=4]{siunitx}
\usepackage{isabelle}
\usepackage{booktabs}
\usepackage[frozencache,cachedir=.]{minted}
\usepackage{tabularx}
\usepackage{multirow}
\usepackage{csquotes}
\usepackage{authblk}
\usepackage{enumitem}
\newunicodechar{∈}{$\in$}

\addto\extrasenglish{

}

\newcommand{\corr}[3]{\ensuremath{}}

\title{A Linter for Isabelle: Implementation and Evaluation}
\author{Yecine Megdiche\thanks{These authors contributed equally to this work.}}
\newcommand\CoAuthorMark{\footnotemark[\arabic{footnote}]}
\author{Fabian Huch\protect\CoAuthorMark}
\author{Lukas Stevens}
\affil{Technische Universität München}
\affil{\texttt{\{megdiche,stevensl,huch\}@in.tum.de}}
\date{May 2022}

\bibliography{references}
\newcommand{\numlints}{22}

\newcommand{\holfoundationallints}{480}
\newcommand{\holsolvedlints}{168}
\newcommand{\distfoundationallints}{14016}
\newcommand{\afpfoundationallints}{59573}
\newcommand{\afpmandatorylints}{1595}

\newcommand{\holperline}{189.997916667}
\newcommand{\distperline}{50.052725457}
\newcommand{\afpperline}{51.175280748}

\newcommand{\holerror}{51.041667}
\newcommand{\holwarn}{46.458333}

\newcommand{\afperror}{55.820590}
\newcommand{\afpwarn}{38.947174}

\begin{document}
\newcolumntype{R}{>{\raggedleft\arraybackslash}X}
\maketitle



\begin{abstract}
In interactive theorem proving, formalization quality is a key factor for maintainability and re-usability of developments and can also impact proof-checking performance.
Commonly, anti-patterns that cause quality issues are known to experienced users.
However, in many theorem prover systems, there are no automatic tools to check for their presence and make less experienced users aware of them.
We attempt to fill this gap in the Isabelle environment by developing a linter as a publicly available add-on component.
The linter offers basic configurability, extensibility, Isabelle/jEdit integration, and a standalone command-line tool.
We uncovered \num{\holfoundationallints} potential problems in Isabelle/HOL,
\num{\distfoundationallints} in other formalizations of the Isabelle distribution,
and an astonishing \num{\afpfoundationallints} in the AFP\@.
With a specific lint bundle for AFP submissions, we found that submission guidelines were violated in \num{\afpmandatorylints} cases.
We set out to alleviate problems in Isabelle/HOL and solved \num{\holsolvedlints} of them so far;
we found that high-severity lints corresponded to actual problems most of the time,
individual users often made the same mistakes in many places,
and that solving those problems retrospectively amounts to a substantial amount of work.
In contrast,
solving these problems interactively for new developments usually incurs only little overhead,
as we found in a quantitative user survey with $22$ participants (less than a minute for more than \SI{60}{\percent} of participants).
We also found that a good explanation of problems is key to the users' ease of solving these problems (correlation coefficient $0.48$), and their satisfaction with the end result (correlation coefficient $0.62$).
\end{abstract}

\section{Introduction}\label{sec:introduction}

Interactive theorem provers make it possible to formalize and verify mathematical concepts in an expressive and interactive way
by constructing machine-verifiable proofs.
However, not all formalizations are created equal:
some may be harder to read,
diverge from standard naming conventions,
have worse run-time characteristics,
or be more prone to break on minor changes.
Consider the following example from HOL-Analysis:
\begin{minted}{isabelle}
apply (simp only: ...)
  apply safe
proof -
  fix i :: 'a
  assume i: "i ∈ Basis"
  ...
  then show "..."
\end{minted}
The proof starts with iterative tactic applications (the \texttt{apply} commands)
and then switches to a structured Isar proof,
where the goals are explicitly stated.
The structured proof is prone to break if the goals generated by the apply-script slightly change,
which may occur, for instance, due to improvements in the simplifier.
It is also hard to read and understand without running Isabelle.
A fully structured proof would be better here,
and would also allow checking individual proof steps in parallel,
whereas the \texttt{apply} steps need to be executed in sequence.

Correctly identifying such problematic constructs
-- especially if they are hidden deep inside a lengthy theory with complicated formalizations --
requires time, effort, and an experienced eye that knows the standards to adhere to.
Hence, searching manually for such problems is not an option,
especially for large theory collections such as the Archive of Formal Proofs (AFP)
which incorporates millions of lines of formalization code.

\textbf{Problem.}
As a result, quality issues can slip into developments unnoticed
and increase the effort needed to perform changes in base sessions
(e.g., when adding new solvers to the simplifier),
make proofs harder to follow and understand, formalizations harder to re-use,
and might cost performance.

\textbf{Solution.}
Static analysis tools that detect anti-patterns and warn about bad practices
-- called \emph{linters} --
are commonly employed to solve this kind of problem in the field of software engineering.
We propose a linter component for Isabelle that gives users feedback during interactive sessions
and as a stand-alone tool for finished theories,
which could be integrated into existing automated build pipelines.

\textbf{Contribution.}
We implemented an Isabelle/Isar linter for Isabelle/HOL as an add-on component\footnote{Isabelle linter: \url{https://github.com/isabelle-prover/isabelle-linter}}
for usage in Isabelle/jEdit as well as through a dedicated command line tool.
It includes $23$ checks, mainly from \emph{Gerwin's Style Guide for Isabelle/HOL}~\cite{klein_2015,klein_2015_2}.
In this paper, we lay out our architecture and implementation,
discuss the uncovered lints quantitatively for the AFP and qualitatively for Isabelle/HOL,
and evaluate the tool in a user study.

\textbf{Organization.}
\autoref{sec:related} covers related tools.
The linter is explained in \autoref{sec:linter};
we evaluate the results in \autoref{sec:evaluation},
and break down the open points in \autoref{sec:future}.

\section{Related Work}\label{sec:related}
The term \emph{lint} originates from the \enquote{Lint} program developed by \citeauthor{johnson1977lint} in~\citeyear{johnson1977lint},
which checks C code \enquote{for bugs and obscurities},
for example by performing more involved type checking than the compilers at the time~\cite{johnson1977lint}.
Nowadays, linters are a standard tool in software engineering,
and implementations exist for most programming languages:
\emph{ESLint} for JavaScript\footnote{ESLint: \url{https://eslint.org/}},
\emph{HLint} for Haskell\footnote{HLint by Neil Mitchell: \url{https://github.com/ndmitchell/hlint}},
and \emph{pylint} for Python\footnote{Pylint - code analysis for Python: \url{https://www.pylint.org/}}, to name a few.
They provide feedback to users to help them catch bugs early, and learn
the best-practices of the respective languages.
Common features of these linters include IDE and CI/CD integration,
and automatic application of the generated suggestions.

For most ITPs,
no linter tools exist yet.
Instead, style guides and user manuals inform users of the best practices.
\emph{Coq}~\cite{barras:inria-00069968}, for example, includes a development style guide on its GitHub repository~\cite{coq_style}.
Projects using Coq, such as \emph{Vericert}~\cite{herklotz2020formal},
frequently provide their own guides on what is expected from code within their source.
The \emph{Agda}~\cite{agda} standard library also includes a style guide highlighting best-practices~\cite{agda_stlib_style}.

For Isabelle/HOL, \citeauthor{klein_2015} published a style guide on their blog~\cite{klein_2015,klein_2015_2},
focusing on readability and maintainability.
They discuss a number of anti-patterns that should be avoided when writing proofs,
the potential problems they induce,
and explains how to avoid them.
The \emph{seL4} project~\cite{seL4} provides some automatic enforcement in form of a \emph{GitHub action}\footnote{GitHub action: \url{https://github.com/seL4/ci-actions/tree/master/thylint}},
which ensures that contributions follow specific rules.
However, it only covers left-over diagnostic and proof-finder commands (which are meant for interactive use) --- a small part of the discussed lints.
There is no integration with any prover IDE\@.

In contrast, for the \emph{Lean} project~\cite{deMoura2015} ---
where much emphasis is put on code quality and user-friendliness ---
a sophisticated \texttt{\#lint} command exists.
In addition to simple syntactic issues
(e.g., unnecessary name-space prefixes, or missing documentation),
the linter also checks for problems in simplification lemmas,
as well as potential mistakes induced by idiosyncrasies of the type class system~\cite{Maintaining2020Doorn}.
However, no detailed quantitative analysis of lint results
and no empirical data on the impact of the linter
(e.g., whether it is helpful to new users)
is available.
\section{Linter Architecture and Implementation}\label{sec:linter}

Our linter analyzes Isabelle theories based on the outer syntax tokens,
i.e., it does not process inner syntax such as terms.
All lints are defined by a name, a severity (info, warning, or error),
and a function that takes the commands of the current document snapshots and a report,
to which it appends its results.
We introduce several lint abstractions to the base interface
to lower the implementation effort for concrete lints.

\subsection{Lint Abstractions}
Lints can conceptually be understood as parsers:
Rather than parsing the grammar of theories, they parse the undesired antipattern.
Hence, we introduced a \emph{parser lint} abstraction that can be used to define lints using parser combinators.
For example, we use it for the \enquote{global attribute on unnamed lemma} lint,
which detects when an unnamed lemma has an attribute such as \texttt{simp} or \texttt{cong}.

Still, it is often easier to implement lints as simple analysis on a partial abstract syntax tree,
which only contains the  elements relevant for linting.
To that end, we implemented an \emph{AST lint} abstraction,
which allows the implementation to only focus on syntax elements that it is concerned with by pattern-matching and overriding the related method.
As a simplified example, the \enquote{tactic proofs} lint that detects usage of tactics (e.g., \texttt{insert} or \texttt{subgoal\_tac})
could be implemented as follows:
\begin{minted}{scala}
  override def lint_method(...) =
    method.info match {
      case Simple_Method(...) =>
        // check if the method is a tactic
      case Combined_Method(left, _, right, _) =>
        // check the left and the right trees
      case _ => None
    }
\end{minted}

For lints concerned with multiple proper commands
(i.e., no white-spaces or comments),
we introduced a \emph{proper commands lint} abstraction,
which provides a filtered list of commands that implementations must match against.
For example, the \emph{low level apply chain} lint detects a long chain of single rule applications
(such as \texttt{simp} or \texttt{rule}),
which could be replaced by automated search methods.

\subsection{Linter Modules}
\begin{figure}[ht]
  \centering
  \begin{tikzpicture}[
  ]
  \begin{scope}[
    every node/.style = {
      draw, rounded corners, rectangle, font=\small, fill=gray!5,
      drop shadow={shadow scale=1.005, shadow xshift=.33ex, shadow yshift=-.33ex}
    }
    ]
  \node (CLI) {CLI};

  \node[below left=1cm and 3.25cm of CLI] (LS) {Lint Store};
  \node[below=1cm of LS] (IL) {<<interface>> Lint};
  \node[below right=1cm and -0.9cm of IL, align=center] (PCL) {Proper\\Commands\\Lint};
  \node[below left=1cm and -0.9cm of IL, align=center] (SCL) {Single\\Command\\Lint};
  \node[below left=1cm and -0.5cm of SCL] (AL) {AST Lint};
  \node[below right=1cm and -0.5cm of SCL] (PL) {Parser Lint};
  \path let
      \p1=(AL.west),
      \p2=(PCL.east)
    in node[below=of AL.south west, anchor=north west, minimum width=\x2-\x1-\pgflinewidth] (Ls) {Lints};

  \node[below=2.5cm of CLI] (L) {Linter};

  \node[right=4.5cm of CLI.north east, anchor=north west] (P) {Plugin};
  \node[below left=1cm and 0cm of P] (LV) {Linter Variable};
  \node[below=of LV] (LD) {Linter Dockable};
  \node[below=of LD] (LO) {Linter Overlay};

  \node at (CLI |- LO) (PR) {Presenter};
  \node[below left=1cm and -0.1cm of PR] (J) {JSON};
  \node[below=of PR] (T) {Text};
  \node[below right=1cm and -0.1cm of PR] (X) {XML};
  \end{scope}

  \begin{scope}[>=open triangle 45]
  \draw[<-] (PCL) -- (PCL |- Ls.north);
  \draw[<-] (AL) -- (AL |- Ls.north);
  \draw[<-] (PL) -- (PL |- Ls.north);
  \draw[<-] (SCL) -- (AL);
  \draw[<-] (SCL) -- (PL);
  \draw[<-] (IL) -- (SCL);
  \draw[<-] (IL) -- (PCL);
  \draw (LS) -- (IL);

  \draw[<-] (PR) -- (J);
  \draw[<-] (PR) -- (T);
  \draw[<-] (PR) -- (X);
  \draw[<-] (PR) -- (LO);
  \end{scope}

  \node[circle,fill,inner sep=0.05cm] at ($(CLI.north) + (0,0.5)$) (CLID) {};
  \draw (CLID) -- (CLI);
  \node[circle,fill,inner sep=0.05cm] at ($(P.north) + (0,0.5)$) (PD) {};
  \draw (PD) -- (P);

  \begin{scope}[>=stealth, draw=red, every node/.style={font=\small}]
  \draw[->] (P) |- (LV);
  \draw[->] (P) |- (LD);
  \draw[->] (P) |- (LO);

  \path ([yshift=0.075cm]LS.east) edge[out=0,in=180,->] node[above, sloped, text=red] {lint selection} ([yshift=0.075cm]CLI.west);
  \path ([yshift=-0.075cm]LS.east) edge[out=0, in=180,<-] ([yshift=-0.075cm]CLI.west);
  \path ([yshift=0.075cm]LS.east) edge[out=0,in=180,->] node[above, pos=0.4, sloped, text=red] {selection} ([yshift=0.075cm]LV.west);
  \path ([yshift=-0.075cm]LS.east) edge[out=0, in=180,<-] ([yshift=-0.075cm]LV.west);

  \path (CLI) edge[red, transform canvas={xshift=-0.075cm}, ->] (L);
  \path (CLI) edge[red, transform canvas={xshift=0.075cm}, <-] node[xshift=-1cm, pos=0.825] {lint report} (L);

  \path ([xshift=0.075cm]PR.north) edge[out=70,in=0,->, to path={(\tikztostart) .. controls +(0,0.25) and +(0,-0.01) .. ([yshift=0.5cm, xshift=0.5cm]\tikztostart) .. controls +(1.5,1) and +(2,0) ..  (\tikztotarget) \tikztonodes}] ([yshift=0.075cm]CLI.east);
  \path ([xshift=-0.075cm]PR.north) edge[out=70,in=0,<-, to path={(\tikztostart) .. controls +(0,0.25) and +(0,-0.01) .. ([yshift=0.5cm, xshift=0.5cm]\tikztostart) .. controls +(1.5,1) and +(2,0) ..  (\tikztotarget) \tikztonodes}] ([yshift=-0.075cm]CLI.east);

  \path ([yshift=0.075cm]L.east) edge[out=0,in=180,->] node[above, sloped, text=red, pos=0.38] {result} ([yshift=0.075cm]LV.west);
  \path ([yshift=-0.075cm]L.east) edge[out=0,in=180,<-] ([yshift=-0.075cm]LV.west);

  \path (LV) edge[red, transform canvas={xshift=-0.1cm}, ->] (LD);
  \path (LV) edge[red, transform canvas={xshift=0.1cm}, <-] (LD);
  \draw[->] (LD) -- node[draw=none,xshift=0.7cm,text=red] {display} (LO);
  \end{scope}

  \begin{pgfonlayer}{background}
    \path[fill=gray!25] ($ (Ls.south west) + (-.25, -.25) $) rectangle ($ (X.north east |- CLI.north east) + (.25, .25) $);
    \path[fill=gray!25] ($ (Ls.south west -| LO.south west) + (-.25, -.25) $) rectangle ($ (P.north east) + (.25, .25) $);
  \end{pgfonlayer}
  \node[anchor=north west, text=black!70] at (Ls.south west |- CLI.north east) {linter base};
  \node[anchor=north west, text=black!70] at (LO.south west |- P.north east) {jedit linter};
\end{tikzpicture}
  \caption{Architecture overview of the linter with Isabelle/jEdit integration.\label{fig:architecture-overview}}
\end{figure}
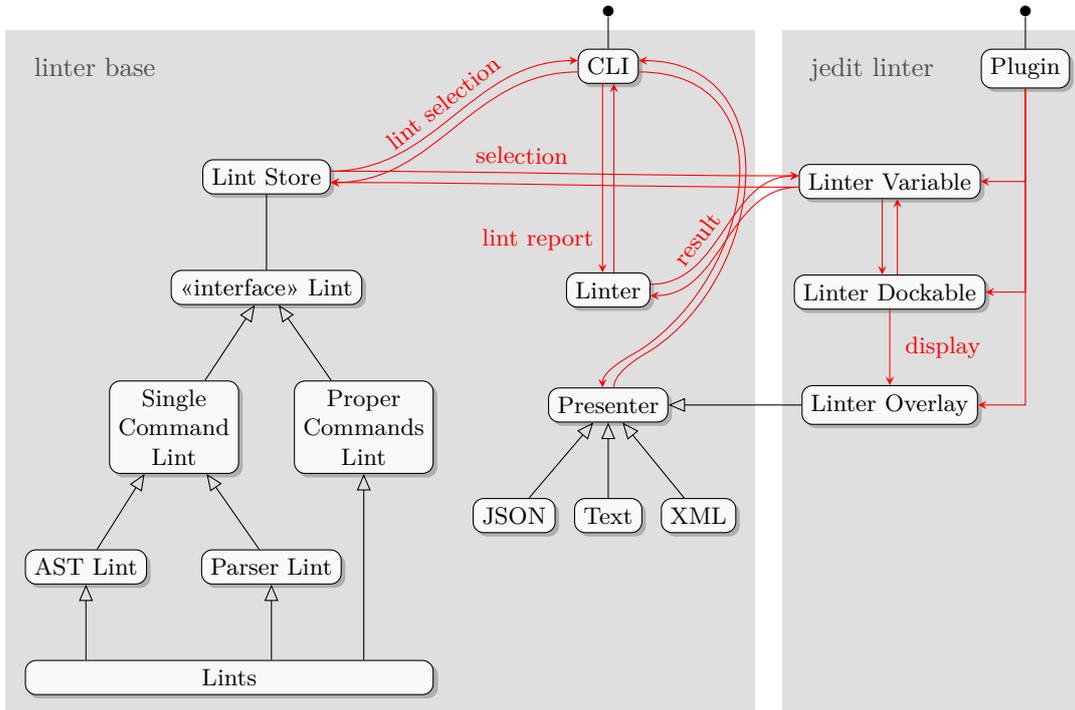

The linter is implemented as Isabelle/Scala module with an additional Isabelle/jEdit plugin.
\autoref{fig:architecture-overview} shows a high-level overview of selected linter components
and their Isabelle interactions, which we discuss in the following.

The linter implementation resides in the \texttt{linter base} package.
It contains a \texttt{Lint Store} module that acts as a repository for lints
and allows registering new lints at runtime,
which permits externally defined lints to be integrated seamlessly.
Also, it introduces the concept of \emph{bundles},
i.e., groups of lints that are used together in a certain context.
Our pre-defined bundles consist of a \emph{foundational} bundle to use when defining object logics or axiomatizations,
the \emph{default} bundle for normal user-space formalizations
(for instance, axiomatizations would trigger lints there),
and a bundle for the \emph{AFP mandatory} guidelines.
Also, there is a \emph{pedantic add-on} bundle for rules that might be considered good practice but are often too tedious to follow,
as well as a \emph{non-interactive add-on} bundle for finished work,
since interactive commands such as \texttt{sledgehammer} are fine while developing proofs interactively,
but usually not for finished sessions.

A selection of lints from the repository
-- which is initialized from linter-specific Isabelle options to control enabled lints and bundles --
can be passed to the actual \texttt{linter} module
to lint a document snapshot.
The result contains triggered lints, positional information, and potentially suggested edits.
Lint results can be transformed into the desired output format with the help of the \texttt{Presenter} module;
implementations are present for a textual format as well as XML and JSON\@.
There are also command-lines modules to run the linter non-interactively
and to generate documentation for lints and bundles.

The Isabelle/jEdit plugin is implemented in the \texttt{jedit linter} package.
It uses the linter via a singleton \texttt{Linter Variable}, which stores the linter configuration 
and caches the results of linting the most recent snapshot of each document,
since the same report might be needed multiple times in different parts of the user interface.

\begin{figure}[htpb]
    \includegraphics[width=\textwidth]{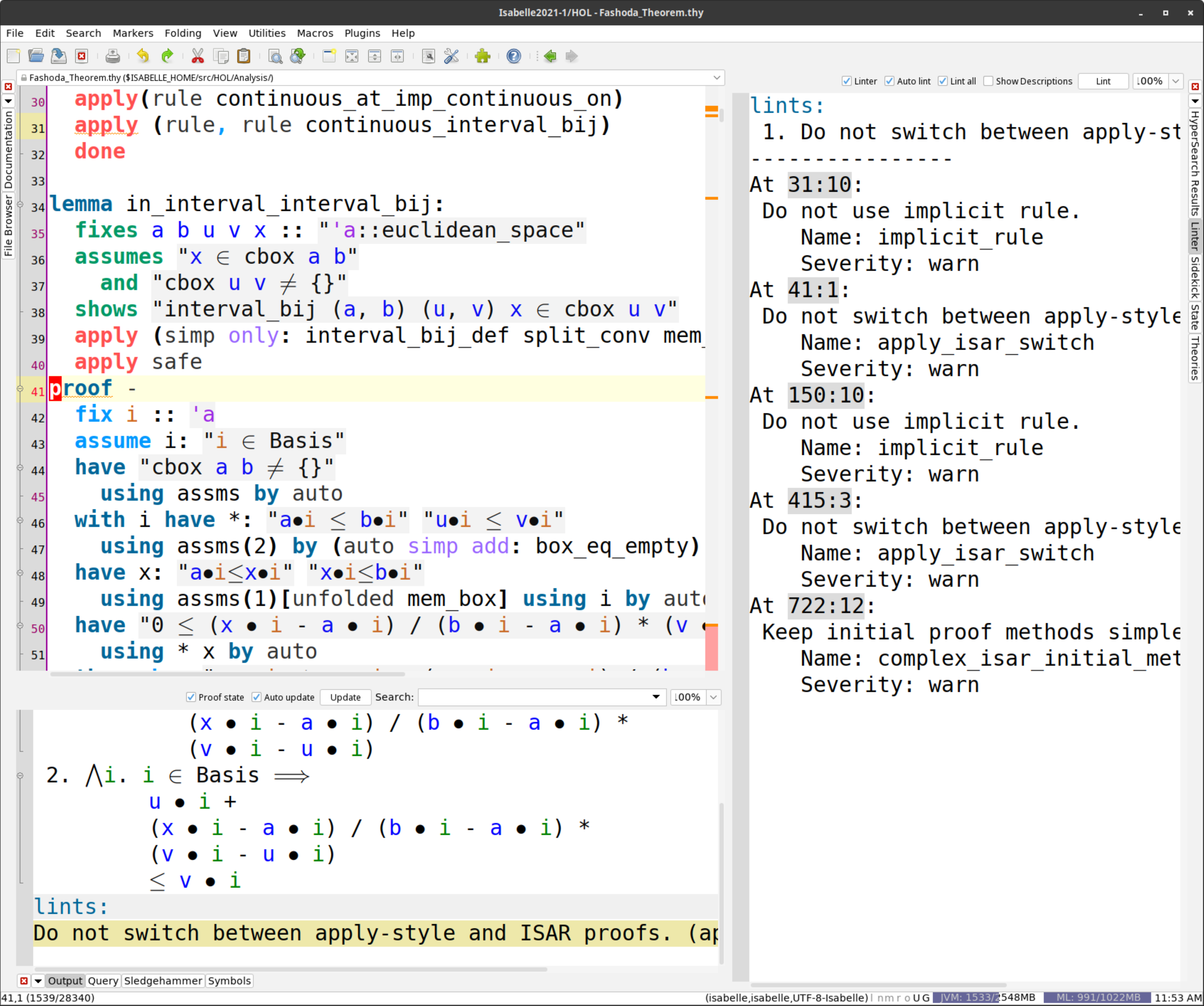}
    \caption{Linter UI elements: Command overlay visible in the buffer and output, linter panel with active position markup.}\label{fig:jedit-buffer}
\end{figure}

As shown in \autoref{fig:jedit-buffer},
lint results are embedded as command overlays (i.e., additional output appended to a command), which are marked in the document buffer and visible in the output panel.
To this end, the \texttt{Linter Overlay} module provides a suitable presenter
and manages the persistent overlay state.
The separate linter panel is implemented in the \texttt{Linter Dockable} module,
which display lints regarding the current command,
an overview over lints of the entire theory,
and control elements for the linter.
We utilize active markup such that one can quickly navigate to lint positions or apply suggestions on click,
similar to other Isabelle commands such as sledgehammer.

\section{Evaluation}\label{sec:evaluation}
We evaluate the linter with the aim to assess the current formalization quality
as well as to find out whether the linter is helpful in guiding users towards better formalizations.
To that end, we formulate our research questions as follows:
\begin{enumerate}[label=\bfseries{RQ\arabic*}:, leftmargin=*]
    \item Are the potential problems understandable for new users?
    \item Do users want to alleviate the potential problems, and are they successful in doing so?
    \item How much did the presence of the linter impact the users' work?
    \item Which problems do we find in the Isabelle distribution and the AFP, and how do they compare?
\end{enumerate}
We illuminated those aspects by different means, which we discuss in the following.

\subsection{User survey}
In order to evaluate the effectiveness of the linter,
we surveyed students of the course \emph{Semantics of Programming Languages} at TUM\@.
Formalization quality was a grading criterion in one of their assignments;
the Linter was recommended but not mandatory to use.
Our survey employed Likert-type questions with five options
(from strong disagreement to strong agreement),
which are mapped to the range $[-2;2]$ in the following.
We report the average of those (avg.), and denote the linear Pearson ($r$) and monotone Spearman ($s$) correlations in APA format (degrees of freedom, correlation coefficient, and significance).

The survey was filled out by $22$ participants,
of which two thirds had some previous Isabelle experience,
most from another lecture using Isabelle.
One fifth had no ITP experience at all,
whereas the rest had experience with another theorem prover.
However, we found that at our significance level of $p=0.05$,
the skill level derived from prior experience did not correlate with the other results.

Most users estimated that they wrote less than one thousand lines with the linter.
When active, lints were reported rarely (\SI{45.5}{\percent}) to occasionally (\SI{36.4}{\percent});
Three participants never experienced a lint, which may be due to misconfiguration\footnote{On  Windows, Isabelle/jEdit needs to be started through the cygwin environment, otherwise the linter is not active.}.

Overall, lint messages communicated the problems quite well
(avg. $1.37$).
Still, it was not as clear to users why the encountered lints could be problems
(avg. $0.53$),
and users did not fully agree that the alleged problems were actual problems (avg. $0.50$).
However, users for whom it was clear why lints could be problems agreed more strongly that they were actual problems
($r(16)=.50$, $p=.034$),
but they also tended to see the problem without the linter
($r(17)=.50$, $p=.019$).

Almost all users discovered some problems that were not apparent without the linter, and changed their code as a result.
For more than \SI{70}{\percent} of users, all change attempts were successful.
Additionally, users who wrote more code with the linter were a bit more successful ($s(16)=.47$, $p=.047$),
i.e., there appears to be a learning factor.

Changes appeared to be mostly easy to perform: over \SI{60}{\percent} took less than a minute, and more than \SI{30}{\percent} between one and ten minutes.
When it was clear why the lints could be problems,
the required effort was significantly less 
($r(16)=-.48$, $p=.046$).
In the end, half of the users were more happy with their resulting code;
one sixth were less happy than before (avg. $0.39$).
Notably, this correlates strongly with whether it was clear why lints could be problems ($r(16)=.62$, $p=.006$)
as well as if users agreed that the lints were actual problems
($r(15)=.60$, $p=.010$).

\subsection{Quantitative Analysis}
To answer RQ4, we compared the results of the linter with our foundational bundle active,
since that is the largest selection of lints applicable in all contexts.
In absolute numbers,
we found a total of \num{\holfoundationallints} lints in Isabelle/HOL,
\num{\distfoundationallints} in the whole distribution,
and \num{\afpfoundationallints} in the AFP (2021-1 releases).
For better comparison, we measured the size of developments by the number of source lines of code
(i.e., without comments and white-space) of their Isabelle theory files.
In Isabelle/HOL, one lint got triggered every \num[round-mode=places,round-precision=1]{\holperline} lines,
compared to every \num[round-mode=places,round-precision=1]{\distperline} lines in the whole distribution
and every \num[round-mode=places,round-precision=1]{\afpperline} lines in the AFP sessions.
While we expected HOL to trigger fewer lints than the other libraries
since it is widely used and kept in shape by many authors,
it is surprising to see that the frequencies of lints in the full Isabelle distribution and the AFP are so close together.

\autoref{tab:lint-summary} shows the number of occurrences per lint as well as severity levels.
It stands out that for HOL and the AFP,
the distribution of severity levels is similar with \SI[round-mode=places,round-precision=1]{\holerror}{\percent}/\SI[round-mode=places,round-precision=1]{\afperror}{\percent} errors and
\SI[round-mode=places,round-precision=1]{\holwarn}{\percent}/\SI[round-mode=places,round-precision=1]{\afpwarn}{\percent} warnings,
but in the Isabelle distribution,
over \SI{70}{\percent} of lints fall into the \enquote{error} category.
Most of them stem from proofs that use tactic methods,
which are problematic since they rely on generated names.
This is likely due to the fact that the distribution contains lots of older material where better methods might not have been around,
whereas the AFP has been growing more recently (doubled in size in the last five years~\cite{Statistics2022Afp}).

\begin{table}[ht]
\centering
\begin{tabularx}{0.9\textwidth}{llRRR}
\toprule
Severity & Lint & HOL & Distribution & AFP\\
\midrule
\multirow[c]{3}{*}{info} & auto\_structural\_composition & \num{2} & \num{15} & \num{264} \\
 & global\_attribute\_changes\cellcolor{gray!10} &\cellcolor{gray!10} \num{0} &\cellcolor{gray!10} \num{15} &\cellcolor{gray!10} \num{52} \\
 & low\_level\_apply\_chain & \num{10} & \num{1051} & \num{2801} \\
\cellcolor{gray!10} & apply\_isar\_switch\cellcolor{gray!10} &\cellcolor{gray!10} \num{1} &\cellcolor{gray!10} \num{54} &\cellcolor{gray!10} \num{1166} \\
\cellcolor{gray!10} & complex\_isar\_initial\_method & \num{127} & \num{848} & \num{4747} \\
\cellcolor{gray!10} & complex\_method\cellcolor{gray!10} &\cellcolor{gray!10} \num{40} &\cellcolor{gray!10} \num{1304} &\cellcolor{gray!10} \num{13526} \\
\cellcolor{gray!10} & implicit\_rule & \num{51} & \num{260} & \num{2786} \\
\cellcolor{gray!10}\multirow[c]{-5}{*}{warn} & lemma\_transforming\_attribute\cellcolor{gray!10} &\cellcolor{gray!10} 4 &\cellcolor{gray!10} \num{530} &\cellcolor{gray!10} \num{977} \\
\multirow[c]{4}{*}{error} & bad\_style\_command & \num{0} & \num{124} & \num{143} \\
 & global\_attribute\_on\_unnamed\_lemma\cellcolor{gray!10} &\cellcolor{gray!10} \num{2} &\cellcolor{gray!10} \num{164} &\cellcolor{gray!10} \num{1010} \\
 & tactic\_proofs & \num{170} & \num{7546} & \num{23215} \\
 & unrestricted\_auto\cellcolor{gray!10} &\cellcolor{gray!10} \num{73} &\cellcolor{gray!10} \num{2105} &\cellcolor{gray!10} \num{8886} \\
\bottomrule
\end{tabularx}
\caption{Absolute number of uncovered lints in Isabelle/HOL, the Isabelle Distribution, and the AFP.}
\label{tab:lint-summary}
\end{table}

While all the lints analyzed previously are problematic for one reason or another,
they are just recommendations for good formalization quality and not obligatory to adhere to.
In contrast, the AFP has some mandatory guidelines for submissions~\cite{Guidelines2022Afp}.
We set up a \emph{afp mandatory} bundle to check for a number of those guidelines automatically\footnote{For technical reasons, the bundle also warns against using the \texttt{apply\_end} keyword,
which is technically not a violation against the submission guidelines and was thus excluded here.}.
In total, the AFP guidelines were violated \num{\afpmandatorylints} times;
\autoref{tab:afp_mandatory} shows how the lints are distributed.
Most violations come from unnamed lemmas that are used globally, for instance when they are added to the simp-set.
However, the AFP submission guidelines aren't fully exhaustive:
For instance,
while they state that counter-example finders such as \texttt{nitpick} must carry an \texttt{expect} attribute
(i.e., \texttt{nitpick} may be used as regression tests),
some entries use nitpick as a model finder through its \texttt{satisfy} option,
which we deem a valid use-case.

\begin{table}[hb]
\centering
\begin{tabular}{lr}
\toprule
Lint & Occurrences \\
\midrule
bad\_style\_command & 117 \\
counter\_example\_finder & 467 \\
global\_attribute\_on\_unnamed\_lemma & 1010 \\
smt\_oracle & 1 \\
\bottomrule
\end{tabular}
\caption{Violations against AFP submission guidelines}
\label{tab:afp_mandatory}
\end{table}

\subsection{Performance}
In a sample from 20 randomly selected sessions of the AFP,
the median time taken to lint a theory
(without generating the build database)
is $20.7$ ms, with a mean of $53.6$ ms.
This means that linting has no noticeable performance impact.
Also, since the linter relies on a normal Isabelle build, that usually needs to be computed anyway,
there is little overhead to using it.

\subsection{Experiences in Isabelle/HOL}
In an effort to improve the formalization quality in Isabelle,
we linted the HOL session and set out to solve the uncovered problems.
To date, we solved \num{\holsolvedlints} of the \num{\holfoundationallints} issues discussed previously.
In doing so,
we confirmed that the severity assigned to the lints
accurately reflect the severity of the reported issues:
for example, the \emph{tactic proofs} lint (severity: error)
generally points to a proof that is hard to read and also hard to update.
The \emph{complex isar initial method} lint (severity: warning) can be solved somewhat automatically\footnote{Although the automatic fix has some improvement potential.}
but represents a serious problem nonetheless.
Finally, the \emph{implicit rule} lint (severity: info) does not hinder readability
and is not particularly problematic.

We also found that individual authors often repeat the same problematic patterns.
This was especially notable by instances of the \emph{complex isar initial method} lint
(where the proof state is transformed with some complex method before starting an Isar proof):
in some theories, structured proofs almost always have a simplifier call in their initial method,
even if not necessary and a simpler \texttt{proof (rule \dots)} or even a standard proof would have been sufficient.
The repetition of anti-patterns indicates that the underlying problem is not known to authors.
Having a linter active and enabled by default
(and configured reasonably)
would address that issue,
and allow users to discover anti-patterns in their work
so they can try to avoid them.

Additionally,
some repetitions were due to historic reasons:
For instance, 
in older theories,
low-level tactics were more prevalent ---
much of the automation that is nowadays available in Isabelle
was not present or less powerful in the early days.

Some of the problems we solved were quite repetitive,
and it became apparent that some lints would benefit from a tighter integration with Isabelle,
for example by utilizing the proof state.
This would allow the linter to automatically resolve some lints,
for instance by finding the rule that was used by an implicit rule application.
In contrast, other issues required re-writing the whole proof
and can thus hardly be solved automatically.
In particular, long apply scripts generally mean that the entire proof needs to be re-written ---
such proofs also were the most tedious to re-write,
but the resulting proofs are much more readable
(and potentially much faster).

A final observation is that most problems did not require a deep understanding of the specific theory at hand:
Isabelle experience, a good grasp of logic and fundamental set theory was sufficient.
Still, familiarity with the underlying concepts is helpful to produce better proofs,
especially where they needed to be re-written completely.

\subsection{Conclusion}
We developed an outer syntax linter for Isabelle/HOL with CLI as well as Isabelle/jEdit integration
and implemented \num{\numlints} checks, mostly from~\cite{klein_2015,klein_2015_2}.
With the help of user survey, performance measurements, a quantitative analysis and our own experiences,
we can now answer our research questions as follows:

\begin{enumerate}[label=\bfseries{RQ\arabic*}:, leftmargin=*]
    \item Even with clearly communicated problems,
new users do not fully understand the underlying problems
and might as a result not see the issue.
    \item Users do utilize the Linter when presented the goal of a high-quality formalization 
and generally set out to solve the linted problems.
Even new users are quite successful in resolving lints in their own formalizations,
though even more so if they understand the underlying problems.
However, it is an order of magnitude harder to solve problems in pre-existing code.
    \item The linter has no discernible performance impact,
and motivated users to quickly solve problems as they arise.
In the end, they were overall happier with the resulting code, which is often much more readable.
    \item In the Isabelle distribution as well as the AFP, more than half of the lints we found were of high severity and are a potential stolperstein for future changes in low-level Isabelle mechanisms.
    The use of tactic methods, calls to \texttt{auto} in the middle of a proof, and overly complex methods were the most frequent violations.
    In terms of lint frequency, the Isabelle distribution and the AFP are very close together.
    Only the HOL session is much better with four times less lints per line of source code.
\end{enumerate}
\section{Future Work}\label{sec:future}
Our linter is readily available and easy to use as an add-on component.
What remains to be done is applying it to existing formalizations
and improve them by solving the uncovered issues.
For that, an integration into existing build pipelines should be built
(e.g., for new AFP submissions)
in order to maintain high quality standards.

The linter itself could be improved with better Isabelle integration, i.e.,
by utilizing theory contexts and linting inner syntax terms.
Moreover, more anti-patterns and conventions could be implemented as lints.
To that end, we ask the Isabelle community to make suggestions on the project page\footnote{Project issues: \url{https://github.com/isabelle-prover/isabelle-linter/issues}}.

Finally, the linter may be integrated into the Isabelle distribution, potentially with support for Isabelle/VSCode over the language server protocol.
\printbibliography{}

\end{document}